\documentclass[%
 preprint,
nofootinbib,
 amsmath,amssymb,
 aps,
prl,
 longbibliography,
 lengthcheck,%
]{revtex4-1}

\usepackage{graphicx}
\usepackage{dcolumn}
\usepackage{bm}
\usepackage{hyperref}

\usepackage{scrextend}
\usepackage{dsfont}

\def\ba{\begin{equation}\begin{array}{c}}
\def\ea{\end{array}\end{equation}}
\def\be{\ba\displaystyle}
\def\ee{\ea}

\newcommand{\tr}{{\rm tr}}

\renewcommand{\Pi}{\hat P}

\newcommand{\const}{{\rm const}}

\begin{document}

\preprint{APS/123-QED}

\title{
Adiabatic theorem for closed quantum systems initialized at finite temperature
}

\author{Nikolai Il`in$^{1}$}
\author{Anastasia Aristova$^{1,3}$}
\author{Oleg Lychkovskiy$^{1,2,3}$}

\affiliation{$^1$ Skolkovo Institute of Science and Technology,
 Bolshoy Boulevard 30, bld. 1, Moscow 121205, Russia}
\affiliation{$^2$ Department of Mathematical Methods for Quantum Technologies, Steklov Mathematical Institute of Russian Academy of Sciences,
8 Gubkina St., Moscow 119991, Russia}
\affiliation{$^3$ Laboratory for the Physics of Complex Quantum Systems, Moscow Institute of Physics and Technology, Institutsky per. 9, Dolgoprudny, Moscow  region,  141700, Russia}


\date{\today}

\begin{abstract}
The evolution of a driven quantum system is said to be adiabatic whenever the state of the system stays close to an instantaneous eigenstate of its time-dependent Hamiltonian. The celebrated quantum adiabatic theorem ensures that such {\it pure state adiabaticity} can be maintained with arbitrary accuracy, provided one chooses a small enough driving rate.  Here, we  extend the notion of quantum adiabaticity to closed quantum systems initially prepared at finite temperature. In this case adiabaticity implies that the (mixed) state of the system stays close to a quasi-Gibbs state diagonal in the basis of the instantaneous eigenstates of the Hamiltonian. We prove a sufficient condition for the finite temperature adiabaticity. Remarkably, it implies that the finite temperature adiabaticity can be more robust than the pure state adiabaticity, particularly in many-body systems. We present an example of a  many-body system where, in the thermodynamic limit, the finite temperature adiabaticity is maintained, while the pure state adiabaticity breaks down.
\end{abstract}

\maketitle


\noindent  {\it Introduction.}~~
A concept of quantum adiabatic evolution was introduced by Born and Fock in the early days of quantum mechanics \cite{Born1926,born1928beweis}.  The concept pertains to a driven closed quantum system described by a time-dependent Hamiltonian. The evolution of the system is called adiabatic as long as the state of the system stays close to the time-dependent instantaneous eigenstate of the Hamiltonian. The celebrated adiabatic theorem \cite{born1928beweis,kato1950} states that adiabaticity can be maintained with any prescribed accuracy, provided the driving rate (i.e. the rate of change of the Hamiltonian) is chosen small enough. The adiabatic theorem enjoys a glorious history and a wide range of theoretical and practical applications, including dynamics of chemical reactions \cite{bowman1991reduced}, population transfer between molecular vibrational levels \cite{gaubatz1990population,bergmann2015perspective},  theory of quantum topological order \cite{budich2013adiabatic}, quantized charge transport \cite{thouless1983quantization}, quantum memory~\cite{fleischhauer2002quantum} and quantum adiabatic computation \cite{farhi2001quantum,albash2018adiabatic,farhi2000quantum}.

Nowadays there is a wealth of experimental techniques available to manipulate large quantum systems consisting of cold atoms in optical lattices, ions in ion traps, arrays of superconducting qubits and quantum dots {\it etc} \cite{2D}. However, these systems are rarely prepared in pure states. Rather, they are typically initialized at some finite temperature determined by the preparation protocol. Therefore the conventional concept of adiabaticity~\cite{Born1926,born1928beweis,kato1950}, which we refer to as {\it pure state adiabaticity} (PSA) in what follows, calls for extension to the case of finite temperature.

Here we define the {\it finite temperature adiabaticity} as the property by which the state of a system initially prepared at finite temperature stays close to the  quasi-Gibbs state in the course of the  unitary quantum evolution. The  time-dependent quasi-Gibbs state, defined by  eq. \eqref{quasi-Gibbs state} below, is  diagonal in the instantaneous eigenbasis of the Hamiltonian and has the same spectrum as the  initial thermal state.

Clearly, if the driving rate is so low that the conditions for PSA for {\it any} eigenstate are met, then the finite temperature adiabaticity is also present, irrespectively of the temperature. It turns out that, in fact, the finite temperature adiabaticity can be present at much higher driving rates. This follows from the finite temperature adiabatic condition proven in the present paper. Remarkably, the energy gaps do not enter this conditions directly, in contrast to the case of PSA. Instead, the role of the energy gaps is played by the temperature. This can be of particular importance for many-body systems, where energy gaps vanish in the thermodynamic limit, and the pure state adiabaticity typically breaks down whenever the driving rate is kept finite but the system size is increased~\cite{FetterWalecka,lychkovskiy2017time}. We provide a particular example of a  many-body system where the finite temperature adiabaticity survives the thermodynamic limit, despite the pure state adiabaticity being broken.


The rest paper is organised as follows. We start from introducing required  definitions and notions (most importantly,  the notion of the quasi-Gibbs state). Then we state the adiabatic theorem for closed quantum systems prepared in thermal states and discuss its scope and implications. After that we illustrate the theorem by applying it to a particular many-body system. We conclude the paper by the summary and outlook. Technical details are relegated to the Supplementary material \cite{supplementary_TAT}.

\medskip

\noindent  {\it Preliminaries \label{sec: prelim}}~~
We describe an isolated driven quantum system by means of a time-dependent Hamiltonian. To introduce time dependence in a way  convenient for our purposes, we consider a Hamiltonian $H_s$ dependent on a parameter $s$ and assume that $s$ varies in time. Without loss of generality, we can assume that $s$ is a linear function of time,
\be
s=\omega t,
\ee
where $\omega$  is the driving rate. The adiabatic limit is defined as
\be\label{adiabatic limit}
\omega\rightarrow0,\qquad t\rightarrow\infty,\qquad \omega t={\rm const}>0.
\ee

Let  $E^n_s$ and $\Phi_{s}^n$ be respectively eigenenergies and eigenvectors of $H_s$,
\be
H_{s} \Phi_{s}^n = E_{s}^n \Phi_{s}^n,\quad n=1,2,...,d,
\ee
where $d$ is the dimension of the Hilbert space. We assume that  $E^n_s$ and $\Phi_{s}^n$ are continuously differentiable in $s$.

Importantly, $H_s$ can be represented as
\be
H_{s}=U_{s}\widetilde{H}_{s}U_{s}^\dag,
\ee
where $U_{s}$ is a continuously differentiable unitary operator,\footnote{Note that $U_{s}$  is {\it not} an evolution operator.}  $U_0=1$, and $\widetilde{H}_s$ is an auxiliary operator  with the same eigenvalues as $H_s$ and the same eigenvectors as $H_0,$
\be
\widetilde{H}_s=\sum_{n}E_{s}^n  \,|n \rangle  \langle n |,
\ee
where $|n \rangle\equiv\Phi_{0}^n$.
Note that time dependence enters $\widetilde{H}_s$ only through $E^n_s$.
An important object in our study is the operator
\be\label{V definition}
 V_s\equiv-iU_{s}^\dag \, \partial_s {U}_{s}.
\ee
To characterize the spectrum, we define
\be\label{mu}
\frac{1}{\mu_s }=\max_{n}\left|\frac{E_0^{n+1}-E_0^n}{E_s^{n+1}-E_s^n}\right|
\ee
and
\be\label{nu}
\nu_s=\max_{n}|\partial_s \ln (E_s^{n+1}-E_s^n)|.
\ee

Often the spectrum of the driven Hamiltonian does not change with time, which we refer to as {\it isospectral driving}. In this case $\widetilde{H}_s=H_0$, $\mu_s$  do not actually depend on $s$, and $\nu_s$ is identically zero. A particular simple instance of the isospectral driving is the {\it uniform} isospectral driving with
\be\label{isospectral driving}
H_s=e^{isV}H_0 \, e^{-isV}.
\ee
Here $V$ coincides with  $V_s$ defined by eq. \eqref{V definition}.

The state of the system $\rho_t$ satisfies the von Neumann equation
\be\label{Schrodiger equation}
i\partial_t \rho_t = [H_{\omega t},\rho_t].
\ee
We assume that at $t=0$ the system is initialized in a thermal state,
\be\label{initial condition}
\rho_0= e^{-\beta H_0}/Z_0,~~~~~~Z_0\equiv \tr \,e^{-\beta H_0},
\ee
$\beta$ being the inverse temperature.

If the system were prepared in an eigenstate  (in particular, in the ground state, i.e. ``at zero temperature''), the  adiabatic theorem~\cite{born1928beweis,kato1950,albash2018adiabatic} would imply that for any given $s$ one can choose sufficiently small $\omega$ so that the state of the system at a (large) time $t=s/\omega$ is close (within a given error margin) to the corresponding instantaneous eigenstate. This is what we refer to as  pure state adiabaticity (PSA).

When we turn to the case of finite temperatures, the first question we have to address is what state one should compare the dynamical state $\rho_t$ with.  If the conditions for PSA are met for any eigenstate, then $\rho_t$  stays close to the {\it quasi-Gibbs} state given by (see also a recent ref. \cite{skelt2020characterizing})
\be\label{quasi-Gibbs state}
\theta_t^\beta \equiv Z_0^{-1}\sum_{n} e^{-\beta E_0^n} |\Phi_{\omega t}^n\rangle\langle \Phi_{\omega t}^n |.
\ee
We will prove that, in fact, this is also the case under different (and, generally, less stringent) conditions that those for PSA.

It should be emphasized that the quasi-Gibbs state \eqref{quasi-Gibbs state} is diagonal in the time-dependent instantaneous eigenbasis of the Hamiltonian, but its spectrum does not change with time and coincides with the spectrum of the initial Gibbs state. The latter feature emerges because the spectrum of the density matrix $\rho_t$ cannot be changed  by the unitary evolution \eqref{Schrodiger equation}.
For this reason the quasi-Gibbs state \eqref{quasi-Gibbs state} is, in general, different from the instantaneous Gibbs state
$
\rho_t^\beta \equiv e^{-\beta H_{\omega t}}/ \tr \,e^{-\beta H_{\omega t}},
$
whose spectrum varies with time.

In what follows we will need to quantify the difference between two mixed quantum states. To this end, we employ the trace distance
\be
D_{\rm tr}(\rho_1,\rho_2)\equiv (1/2)\,\tr |\rho_2-\rho_1 |,
\ee
which is known to have a straightforward operational meaning \cite{helstrom1969quantum,holevo1972quasiequivalence,holevo1973statistical,wilde2013quantum}.



\medskip

\noindent  {\it Adiabatic theorem for finite temperatures. \label{sec: AT}}~~Now we are in a position to state the following

\smallskip

\begin{addmargin}[1em]{0em}
{\bf Theorem:} The trace distance between the dynamical state of the system $\rho_t$ (initialized in the Gibbs state \eqref{initial condition} and evolving according to the von Neumann equation \eqref{Schrodiger equation}) and the quasi-Gibbs state $\theta_t^\beta$ (defined by eq. \eqref{quasi-Gibbs state}) is bounded from above by
\begin{eqnarray}\label{main result}
\nonumber
&&D_{\rm tr}\left(\rho_t,\theta_t^\beta\right)\leq \sqrt{\sqrt{2}\omega\beta} \,\Bigg(\frac{1}{\mu_{\omega t}}\|V_{\omega t}\|
\\
\nonumber
&+&\int_{0}^{\omega t}\frac{1}{\mu_{s'}}\|\partial_{ s'}V_{ s'}\|d s'+\int_{0}^{\omega t}\frac{\nu_{s'}}{\mu_{s'}}\|V_{ s'}\|d s'
\\
&+&
\sqrt2 \, \int_{0}^{\omega t}\frac{1}{\mu_{s'}}\|V_{ s'}\|^2d s'\Bigg)^{1/2}.
\end{eqnarray}
Here $V_s$, $\mu_s$ and $\nu_s$ are defined according to eqs. \eqref{V definition}, \eqref{mu} and \eqref{nu}, respectively, and $\|\dots\|$ refers to the operator norm.\footnote{For our purposes, the operator norm $\|\dots\|$ can be defined  as the maximum among absolute values of eigenvalues of the corresponding operator.}
\end{addmargin}

\smallskip

\noindent This theorem implies that $\rho_t$ converges to $\theta_t^\beta$  in the adiabatic limit \eqref{adiabatic limit}, provided the term in brackets remains finite.  The proof of the theorem can be found in the Supplement~\cite{supplementary_TAT}.

Observe that the r.h.s. of the bound \eqref{main result} vanishes in the limit of infinite temperature, $\beta=0$. This is consistent with the simple fact that at the infinite temperature $\rho_t=\theta_t^{\beta=0}=\mathds{1}/d$, and the evolution is adiabatic at any driving rate.

The theorem admits a particularly simple form in the case of the uniform isospectral driving \eqref{isospectral driving}:

\smallskip

\begin{addmargin}[0em]{0em}
{\bf Corollary:} For the isospectrally and uniformly driven Hamiltonian \eqref{isospectral driving} the bound \eqref{main result} reads
\begin{equation}\label{corollary}
D_{\rm tr}\left(\rho_t,\theta_t^\beta\right)\leq \sqrt{\sqrt{2}\omega\beta \,\|V\|\,\left(1+
\sqrt2 \, \omega t\,\|V\| \right)}.
\end{equation}
\end{addmargin}

\smallskip

The corollary immediately implies that $\rho_t$ converges to $\theta_t^\beta$  in the adiabatic limit~\eqref{adiabatic limit} whenever $\|V\|$ is finite.

Remarkably, energy gaps do not directly enter the bounds \eqref{main result} and \eqref{corollary},
in contrast to typical sufficient conditions for PSA \cite{albash2018adiabatic} (see, however, \cite{avron1999adiabatic,teufel2001note}). This is crucial for the robustness of adiabaticity in the thermodynamic limit, since the energy gaps vanish with increasing the system size. The system size may also enter the bounds \eqref{main result} and \eqref{corollary} through  $\|V_s\|$, $\|\partial_s V_s\|$ and (for the bound \eqref{main result}) through $\mu_s$, $\nu_s$. When the above quantities are finite in the thermodynamic limit, the finite temperature adiabaticity  survives in this limit even if the PSA fails.  Below we consider a  many-body system exhibiting such behavior.


\begin{figure}
\begin{center}
\includegraphics[width=\linewidth]{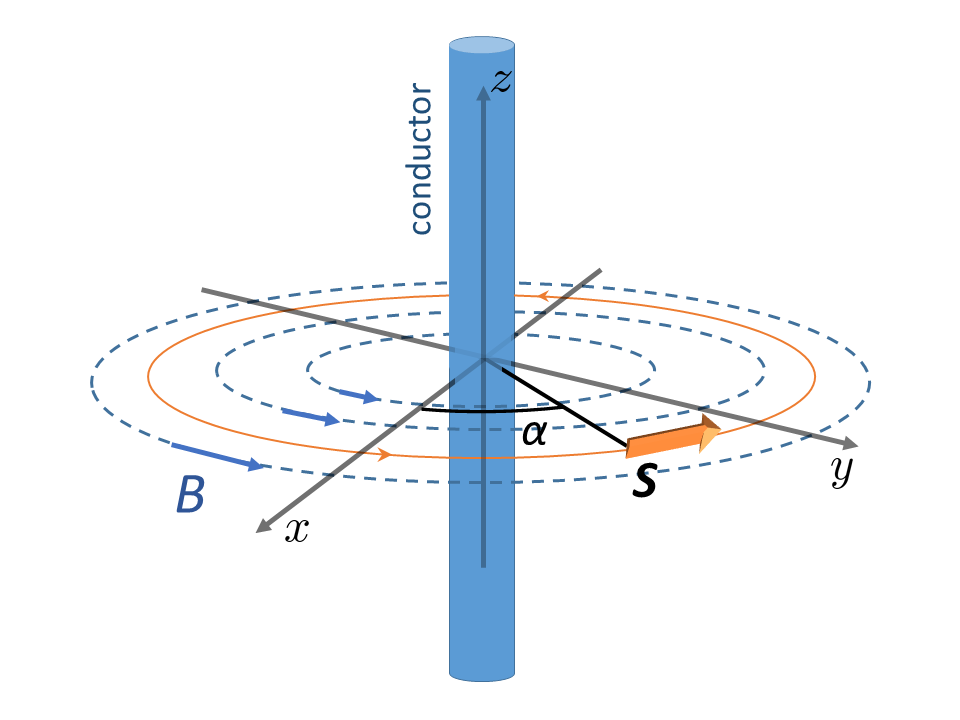}
\end{center}
\caption{(Color online) A quantum sensor with a single spin possessing a magnetic moment is  moved around a wire along a circular trajectory. The net current through the wire is zero, however the electrons in the wire are still magnetically coupled to the spin due to fluctuations of the current, see eqs. \eqref{H wire}, \eqref{H Se}. The many-body adiabaticity of the electron-spin system  at finite temperature is robust with respect to increasing the system size (i.e. the length of the wire).  In contract, the pure state adiabaticity breaks down in the thermodynamic limit at any finite driving rate.}
\label{fig}
\end{figure}

\medskip

\noindent  {\it Example.}~~ Consider a thin straight wire with $N$ electrons and a quantum sensor which can be moved around the wire, see Fig. \ref{fig}. We consider a toy model of the sensor consisting of a single quantum spin $S$ with a magnetic moment $\mu_{\rm magn}$ (not to be confused with $\mu_s$ defined in eq. \eqref{mu}). The interaction between the spin and the electrons is mediated by the magnetic field produced by the electron motion.\footnote{We disregard the magnetic fields of the magnetic moments of electrons.} We consider the case of zero net current of electrons. Still, the interaction persists even in this case  due to fluctuations of the current, both classical and quantum. The Hamiltonian of the system reads
\begin{align}\label{H wire}
H_\alpha= & \,H^e+H^{Se}_\alpha,
\end{align}
where $H_e$ is the Hamiltonian of electrons (we do not need its explicit form here), and
\begin{align}\label{H Se}
H^{Se}_\alpha=  & \,-\frac{\mu_{\rm magn}}{2\pi r} \, {\cal J} \left(-\sin\alpha\, S_x+\cos\alpha \, S_y\right)
\end{align}
is the Hamiltonian of the magnetic field-mediated interaction between electrons and the spin. Here  $(S_x,S_y,S_z)$ are the components of the spin operator, ${\cal J}$ is the operator of the electron current, $r$ is the distance from the sensor to the wire and $\alpha$ is the polar angle determining the position of the sensor, see Fig. \ref{fig}.

We further assume that the sensor is moved along a circular trajectory around the wire with $r=\const$ and $\alpha = \omega t$. Then the Hamiltonian \eqref{H wire} can be cast in the form~\eqref{isospectral driving}, $H_\alpha=e^{-i\alpha S_z} H_{\alpha=0} \,e^{i\alpha S_z}$, therefore the bound~\eqref{corollary} with $V=-S_z$ applies. This bound implies that
 it suffices to choose
\be\label{wire sufficient condition}
\omega\leq \frac{\varepsilon^2}{\sqrt2 \, \beta \, S\,(1+\sqrt2 \,\alpha\, S)}
\ee
to move the sensor up to the angle $\alpha$ along the circular  trajectory while maintaining adiabaticity with precision $\varepsilon$, $\left.D_{\rm tr}\left(\rho_{t},\theta_t^\beta\right)\right|_{t=\alpha/\omega}\leq\varepsilon$.

Remarkably, the sufficient adiabatic condition \eqref{wire sufficient condition} does not depend on the number of electrons. Thus the finite temperature adiabaticity is robust in the thermodynamic limit $N\rightarrow\infty$, $L\rightarrow\infty$, $A=\const$, $\rho\equiv N/(LA)=\const$, where $L$ and $A$ are, respectively, the length and the cross section of the wire, and $\rho$ is the number density of electrons in the wire.

In contract, the pure state adiabaticity breaks down in the thermodynamic limit. This can be easily seen if periodic boundary conditions along the $z$ direction are imposed on the electron wave functions. In this case the Hamiltonian~\eqref{H wire} commutes with the current operator, the latter being related to the total momentum of electrons, $P_e$,
\be\label{current}
{\cal J} =\frac{e\rho A }{m_e N} \,P_e,
\ee
where $e$ and  $m_e$ are the charge and the mass of the electron. As a result, the dynamics of the spin is governed by the effective Hamiltonian \eqref{H Se}, where ${\cal J}$ now refers to the eigenvalue of the current operator \eqref{current} in the eigenstate the system is initialized in. Since in the typical eigenstate from the Gibbs ensemble this eigenvalue is $O(1/\sqrt N)$, the driving rate necessary to maintain the pure state adiabaticity also scales as $1/\sqrt N$ and vanishes in the thermodynamic limit (see a detailed analysis in the Supplement \cite{supplementary_TAT}).

\medskip

\noindent  {\it Summary and outlook \label{sec: conclusions}}

To summarize, we have introduced the notion of finite temperature adiabaticity of an isolated quantum system and proved the finite temperature adiabatic theorem \eqref{main result}. The sufficient adiabatic condition which follows from this theorem does not contain energy gaps, in contrast to most of the adiabatic conditions for pure state adiabaticity. This indicates that the finite temperature adiabaticity can be more robust in the thermodynamic limit then the pure state adiabaticity. We confirm this expectation for the specific model~\eqref{H wire}. It should be noted that this robustness is consistent with earlier numerical observations that microcanonical mixed states are more robust to adiabaticity breaking than pure states \cite{burkle2020probabilistic}.

It should be emphasized that our notion of adiabaticity refers to the many-body state of the system and is different from the notion of local adiabaticity  \cite{abou-salem2005adiabatic,abou-salem2007status,jaksic2014note,bachmann2016adiabatic,venuti2016adiabaticity,benoist2016full,teufel2019non-equilibrium}. The latter notion applies to the reduced density matrix of a subsystem coupled to a reservoir.  The many-body adiabaticity implies the local adiabaticity, but not vice versa.

A considerable limitation of the bounds \eqref{main result}, \eqref{corollary} is that they contain the operator norms. For continuous systems operator norms of certain physically relevant operators (e.g. momentum) are infinite, which renders the bounds void. In fact, the operator norm can be replaced by the better behaved thermal averages in {\it some} of the terms in eqs. \eqref{main result}, \eqref{corollary}, as we discuss in the Supplement \cite{supplementary_TAT}. However, at the moment we are not able to avoid the operator norms altogether, and leave the improvement of the bounds  \eqref{main result}, \eqref{corollary} in this direction  for further work.

\smallskip

\begin{acknowledgements}
\noindent  {\it Acknowledgements.} We are grateful to V. Dobrovitski for a useful discussion. The work was supported by the Russian Science Foundation under the grant N$^{\rm o}$ 17-71-20158.
\end{acknowledgements}

\bibliography{C:/D/Work/QM/Bibs/1D,C:/D/Work/QM/Bibs/LZ_and_adiabaticity,C:/D/Work/QM/Bibs/AQC,C:/D/Work/QM/Bibs/QIP}

\providecommand{\noopsort}[1]{}\providecommand{\singleletter}[1]{#1}
\begin{thebibliography}{31}%
\makeatletter
\providecommand \@ifxundefined [1]{%
 \@ifx{#1\undefined}
}%
\providecommand \@ifnum [1]{%
 \ifnum #1\expandafter \@firstoftwo
 \else \expandafter \@secondoftwo
 \fi
}%
\providecommand \@ifx [1]{%
 \ifx #1\expandafter \@firstoftwo
 \else \expandafter \@secondoftwo
 \fi
}%
\providecommand \natexlab [1]{#1}%
\providecommand \enquote  [1]{``#1''}%
\providecommand \bibnamefont  [1]{#1}%
\providecommand \bibfnamefont [1]{#1}%
\providecommand \citenamefont [1]{#1}%
\providecommand \href@noop [0]{\@secondoftwo}%
\providecommand \href [0]{\begingroup \@sanitize@url \@href}%
\providecommand \@href[1]{\@@startlink{#1}\@@href}%
\providecommand \@@href[1]{\endgroup#1\@@endlink}%
\providecommand \@sanitize@url [0]{\catcode `\\12\catcode `\$12\catcode
  `\&12\catcode `\#12\catcode `\^12\catcode `\_12\catcode `\%12\relax}%
\providecommand \@@startlink[1]{}%
\providecommand \@@endlink[0]{}%
\providecommand \url  [0]{\begingroup\@sanitize@url \@url }%
\providecommand \@url [1]{\endgroup\@href {#1}{\urlprefix }}%
\providecommand \urlprefix  [0]{URL }%
\providecommand \Eprint [0]{\href }%
\providecommand \doibase [0]{http://dx.doi.org/}%
\providecommand \selectlanguage [0]{\@gobble}%
\providecommand \bibinfo  [0]{\@secondoftwo}%
\providecommand \bibfield  [0]{\@secondoftwo}%
\providecommand \translation [1]{[#1]}%
\providecommand \BibitemOpen [0]{}%
\providecommand \bibitemStop [0]{}%
\providecommand \bibitemNoStop [0]{.\EOS\space}%
\providecommand \EOS [0]{\spacefactor3000\relax}%
\providecommand \BibitemShut  [1]{\csname bibitem#1\endcsname}%
\let\auto@bib@innerbib\@empty
\bibitem [{\citenamefont {Born}(1926)}]{Born1926}%
  \BibitemOpen
  \bibfield  {author} {\bibinfo {author} {\bibfnamefont {Max}\ \bibnamefont
  {Born}},\ }\bibfield  {title} {\enquote {\bibinfo {title} {Das
  adiabatenprinzip in der quantenmeehanik},}\ }\href@noop {} {\bibfield
  {journal} {\bibinfo  {journal} {Zeitschrift f{\"u}r Physik}\ }\textbf
  {\bibinfo {volume} {40}},\ \bibinfo {pages} {167} (\bibinfo {year}
  {1926})}\BibitemShut {NoStop}%
\bibitem [{\citenamefont {Born}\ and\ \citenamefont
  {Fock}(1928)}]{born1928beweis}%
  \BibitemOpen
  \bibfield  {author} {\bibinfo {author} {\bibfnamefont {Max}\ \bibnamefont
  {Born}}\ and\ \bibinfo {author} {\bibfnamefont {Vladimir}\ \bibnamefont
  {Fock}},\ }\bibfield  {title} {\enquote {\bibinfo {title} {Beweis des
  adiabatensatzes},}\ }\href@noop {} {\bibfield  {journal} {\bibinfo  {journal}
  {Zeitschrift f{\"u}r Physik}\ }\textbf {\bibinfo {volume} {51}},\ \bibinfo
  {pages} {165--180} (\bibinfo {year} {1928})}\BibitemShut {NoStop}%
\bibitem [{\citenamefont {Kato}(1950)}]{kato1950}%
  \BibitemOpen
  \bibfield  {author} {\bibinfo {author} {\bibfnamefont {Tosio}\ \bibnamefont
  {Kato}},\ }\bibfield  {title} {\enquote {\bibinfo {title} {On the adiabatic
  theorem of quantum mechanics},}\ }\href {\doibase 10.1143/JPSJ.5.435}
  {\bibfield  {journal} {\bibinfo  {journal} {Journal of the Physical Society
  of Japan}\ }\textbf {\bibinfo {volume} {5}},\ \bibinfo {pages} {435--439}
  (\bibinfo {year} {1950})}\BibitemShut {NoStop}%
\bibitem [{\citenamefont {Bowman}(1991)}]{bowman1991reduced}%
  \BibitemOpen
  \bibfield  {author} {\bibinfo {author} {\bibfnamefont {Joel~M}\ \bibnamefont
  {Bowman}},\ }\bibfield  {title} {\enquote {\bibinfo {title} {Reduced
  dimensionality theory of quantum reactive scattering},}\ }\href@noop {}
  {\bibfield  {journal} {\bibinfo  {journal} {The Journal of Physical
  Chemistry}\ }\textbf {\bibinfo {volume} {95}},\ \bibinfo {pages} {4960--4968}
  (\bibinfo {year} {1991})}\BibitemShut {NoStop}%
\bibitem [{\citenamefont {Gaubatz}\ \emph {et~al.}(1990)\citenamefont
  {Gaubatz}, \citenamefont {Rudecki}, \citenamefont {Schiemann},\ and\
  \citenamefont {Bergmann}}]{gaubatz1990population}%
  \BibitemOpen
  \bibfield  {author} {\bibinfo {author} {\bibfnamefont {U}~\bibnamefont
  {Gaubatz}}, \bibinfo {author} {\bibfnamefont {P}~\bibnamefont {Rudecki}},
  \bibinfo {author} {\bibfnamefont {S}~\bibnamefont {Schiemann}}, \ and\
  \bibinfo {author} {\bibfnamefont {K}~\bibnamefont {Bergmann}},\ }\bibfield
  {title} {\enquote {\bibinfo {title} {Population transfer between molecular
  vibrational levels by stimulated raman scattering with partially overlapping
  laser fields. a new concept and experimental results},}\ }\href@noop {}
  {\bibfield  {journal} {\bibinfo  {journal} {The Journal of Chemical Physics}\
  }\textbf {\bibinfo {volume} {92}},\ \bibinfo {pages} {5363--5376} (\bibinfo
  {year} {1990})}\BibitemShut {NoStop}%
\bibitem [{\citenamefont {Bergmann}\ \emph {et~al.}(2015)\citenamefont
  {Bergmann}, \citenamefont {Vitanov},\ and\ \citenamefont
  {Shore}}]{bergmann2015perspective}%
  \BibitemOpen
  \bibfield  {author} {\bibinfo {author} {\bibfnamefont {Klaas}\ \bibnamefont
  {Bergmann}}, \bibinfo {author} {\bibfnamefont {Nikolay~V}\ \bibnamefont
  {Vitanov}}, \ and\ \bibinfo {author} {\bibfnamefont {Bruce~W}\ \bibnamefont
  {Shore}},\ }\bibfield  {title} {\enquote {\bibinfo {title} {Perspective:
  Stimulated raman adiabatic passage: The status after 25 years},}\ }\href@noop
  {} {\bibfield  {journal} {\bibinfo  {journal} {The Journal of chemical
  physics}\ }\textbf {\bibinfo {volume} {142}},\ \bibinfo {pages} {170901}
  (\bibinfo {year} {2015})}\BibitemShut {NoStop}%
\bibitem [{\citenamefont {Budich}\ and\ \citenamefont
  {Trauzettel}(2013)}]{budich2013adiabatic}%
  \BibitemOpen
  \bibfield  {author} {\bibinfo {author} {\bibfnamefont {Jan~Carl}\
  \bibnamefont {Budich}}\ and\ \bibinfo {author} {\bibfnamefont {Bj{\"o}rn}\
  \bibnamefont {Trauzettel}},\ }\bibfield  {title} {\enquote {\bibinfo {title}
  {From the adiabatic theorem of quantum mechanics to topological states of
  matter},}\ }\href@noop {} {\bibfield  {journal} {\bibinfo  {journal} {physica
  status solidi (RRL)-Rapid Research Letters}\ }\textbf {\bibinfo {volume}
  {7}},\ \bibinfo {pages} {109--129} (\bibinfo {year} {2013})}\BibitemShut
  {NoStop}%
\bibitem [{\citenamefont {Thouless}(1983)}]{thouless1983quantization}%
  \BibitemOpen
  \bibfield  {author} {\bibinfo {author} {\bibfnamefont {DJ}~\bibnamefont
  {Thouless}},\ }\bibfield  {title} {\enquote {\bibinfo {title} {Quantization
  of particle transport},}\ }\href@noop {} {\bibfield  {journal} {\bibinfo
  {journal} {Physical Review B}\ }\textbf {\bibinfo {volume} {27}},\ \bibinfo
  {pages} {6083} (\bibinfo {year} {1983})}\BibitemShut {NoStop}%
\bibitem [{\citenamefont {Fleischhauer}\ and\ \citenamefont
  {Lukin}(2002)}]{fleischhauer2002quantum}%
  \BibitemOpen
  \bibfield  {author} {\bibinfo {author} {\bibfnamefont {M.}~\bibnamefont
  {Fleischhauer}}\ and\ \bibinfo {author} {\bibfnamefont {M.~D.}\ \bibnamefont
  {Lukin}},\ }\bibfield  {title} {\enquote {\bibinfo {title} {Quantum memory
  for photons: Dark-state polaritons},}\ }\href {\doibase
  10.1103/PhysRevA.65.022314} {\bibfield  {journal} {\bibinfo  {journal} {Phys.
  Rev. A}\ }\textbf {\bibinfo {volume} {65}},\ \bibinfo {pages} {022314}
  (\bibinfo {year} {2002})}\BibitemShut {NoStop}%
\bibitem [{\citenamefont {Farhi}\ \emph {et~al.}(2001)\citenamefont {Farhi},
  \citenamefont {Goldstone}, \citenamefont {Gutmann}, \citenamefont {Lapan},
  \citenamefont {Lundgren},\ and\ \citenamefont {Preda}}]{farhi2001quantum}%
  \BibitemOpen
  \bibfield  {author} {\bibinfo {author} {\bibfnamefont {E.}~\bibnamefont
  {Farhi}}, \bibinfo {author} {\bibfnamefont {J.}~\bibnamefont {Goldstone}},
  \bibinfo {author} {\bibfnamefont {S.}~\bibnamefont {Gutmann}}, \bibinfo
  {author} {\bibfnamefont {J.}~\bibnamefont {Lapan}}, \bibinfo {author}
  {\bibfnamefont {A.}~\bibnamefont {Lundgren}}, \ and\ \bibinfo {author}
  {\bibfnamefont {D.}~\bibnamefont {Preda}},\ }\bibfield  {title} {\enquote
  {\bibinfo {title} {A quantum adiabatic evolution algorithm applied to random
  instances of an np-complete problem},}\ }\href {\doibase
  10.1126/science.1057726} {\bibfield  {journal} {\bibinfo  {journal}
  {Science}\ }\textbf {\bibinfo {volume} {292}},\ \bibinfo {pages} {472}
  (\bibinfo {year} {2001})}\BibitemShut {NoStop}%
\bibitem [{\citenamefont {Albash}\ and\ \citenamefont
  {Lidar}(2018)}]{albash2018adiabatic}%
  \BibitemOpen
  \bibfield  {author} {\bibinfo {author} {\bibfnamefont {Tameem}\ \bibnamefont
  {Albash}}\ and\ \bibinfo {author} {\bibfnamefont {Daniel~A.}\ \bibnamefont
  {Lidar}},\ }\bibfield  {title} {\enquote {\bibinfo {title} {Adiabatic quantum
  computation},}\ }\href {\doibase 10.1103/RevModPhys.90.015002} {\bibfield
  {journal} {\bibinfo  {journal} {Rev. Mod. Phys.}\ }\textbf {\bibinfo {volume}
  {90}},\ \bibinfo {pages} {015002} (\bibinfo {year} {2018})}\BibitemShut
  {NoStop}%
\bibitem [{\citenamefont {Farhi}\ \emph {et~al.}()\citenamefont {Farhi},
  \citenamefont {Goldstone}, \citenamefont {Gutmann},\ and\ \citenamefont
  {Sipser}}]{farhi2000quantum}%
  \BibitemOpen
  \bibfield  {author} {\bibinfo {author} {\bibfnamefont {E.}~\bibnamefont
  {Farhi}}, \bibinfo {author} {\bibfnamefont {J.}~\bibnamefont {Goldstone}},
  \bibinfo {author} {\bibfnamefont {S.}~\bibnamefont {Gutmann}}, \ and\
  \bibinfo {author} {\bibfnamefont {M.}~\bibnamefont {Sipser}},\ }\bibfield
  {title} {\enquote {\bibinfo {title} {Quantum computation by adiabatic
  evolution},}\ }\href@noop {} {\ }\Eprint
  {http://arxiv.org/abs/arXiv:quant-ph/0001106} {arXiv:quant-ph/0001106}
  \BibitemShut {NoStop}%
\bibitem [{2D()}]{2D}%
  \BibitemOpen
  \href@noop {} {\emph {\bibinfo {title} {2D Quantum Metamaterials: Proceedings
  of the 2018 NIST Workshop)}}}\BibitemShut {NoStop}%
\bibitem [{\citenamefont {Alexander L.~Fetter}(2003)}]{FetterWalecka}%
  \BibitemOpen
  \bibfield  {author} {\bibinfo {author} {\bibfnamefont {John Dirk~Walecka}\
  \bibnamefont {Alexander L.~Fetter}},\ }\href
  {http://gen.lib.rus.ec/book/index.php?md5=F701CEC9C051A4A02C9A883ED45650D5}
  {\emph {\bibinfo {title} {Quantum Theory of Many-Particle Systems}}},\ Dover
  Books on Physics\ (\bibinfo  {publisher} {Dover Publications},\ \bibinfo
  {year} {2003})\BibitemShut {NoStop}%
\bibitem [{\citenamefont {Lychkovskiy}\ \emph {et~al.}(2017)\citenamefont
  {Lychkovskiy}, \citenamefont {Gamayun},\ and\ \citenamefont
  {Cheianov}}]{lychkovskiy2017time}%
  \BibitemOpen
  \bibfield  {author} {\bibinfo {author} {\bibfnamefont {Oleg}\ \bibnamefont
  {Lychkovskiy}}, \bibinfo {author} {\bibfnamefont {Oleksandr}\ \bibnamefont
  {Gamayun}}, \ and\ \bibinfo {author} {\bibfnamefont {Vadim}\ \bibnamefont
  {Cheianov}},\ }\bibfield  {title} {\enquote {\bibinfo {title} {Time scale for
  adiabaticity breakdown in driven many-body systems and orthogonality
  catastrophe},}\ }\href {\doibase 10.1103/PhysRevLett.119.200401} {\bibfield
  {journal} {\bibinfo  {journal} {{P}hys. {R}ev. {L}ett.}\ }\textbf {\bibinfo
  {volume} {119}},\ \bibinfo {pages} {200401} (\bibinfo {year}
  {2017})}\BibitemShut {NoStop}%
\bibitem [{sup()}]{supplementary_TAT}%
  \BibitemOpen
  \href@noop {} {}\bibinfo {note} {See the supplementary material to this
  article for the proof of the finite temperature adiabatic theorem and
  analysis of the pure state adiabaticity in the spin-electron
  model}\BibitemShut {NoStop}%
\bibitem [{\citenamefont {Skelt}\ and\ \citenamefont
  {D'Amico}(2020)}]{skelt2020characterizing}%
  \BibitemOpen
  \bibfield  {author} {\bibinfo {author} {\bibfnamefont {AH}~\bibnamefont
  {Skelt}}\ and\ \bibinfo {author} {\bibfnamefont {I}~\bibnamefont {D'Amico}},\
  }\bibfield  {title} {\enquote {\bibinfo {title} {Characterizing adiabaticity
  in quantum many-body systems at finite temperature},}\ }\href@noop {}
  {\bibfield  {journal} {\bibinfo  {journal} {arXiv:2004.05842}\ } (\bibinfo
  {year} {2020})}\BibitemShut {NoStop}%
\bibitem [{\citenamefont {Helstrom}(1969)}]{helstrom1969quantum}%
  \BibitemOpen
  \bibfield  {author} {\bibinfo {author} {\bibfnamefont {Carl~W}\ \bibnamefont
  {Helstrom}},\ }\bibfield  {title} {\enquote {\bibinfo {title} {Quantum
  detection and estimation theory},}\ }\href@noop {} {\bibfield  {journal}
  {\bibinfo  {journal} {Journal of Statistical Physics}\ }\textbf {\bibinfo
  {volume} {1}},\ \bibinfo {pages} {231--252} (\bibinfo {year}
  {1969})}\BibitemShut {NoStop}%
\bibitem [{\citenamefont {Holevo}(1972)}]{holevo1972quasiequivalence}%
  \BibitemOpen
  \bibfield  {author} {\bibinfo {author} {\bibfnamefont {A~S}\ \bibnamefont
  {Holevo}},\ }\bibfield  {title} {\enquote {\bibinfo {title} {On
  quasiequivalence of locally normal states},}\ }\href {\doibase
  10.1007/BF01035528} {\bibfield  {journal} {\bibinfo  {journal} {Theor. Math.
  Phys.}\ }\textbf {\bibinfo {volume} {13}},\ \bibinfo {pages} {1071--1082}
  (\bibinfo {year} {1972})}\BibitemShut {NoStop}%
\bibitem [{\citenamefont {Holevo}(1973)}]{holevo1973statistical}%
  \BibitemOpen
  \bibfield  {author} {\bibinfo {author} {\bibfnamefont {A~S}\ \bibnamefont
  {Holevo}},\ }\bibfield  {title} {\enquote {\bibinfo {title} {Statistical
  decision theory for quantum systems},}\ }\href {\doibase
  10.1016/0047-259X(73)90028-6} {\bibfield  {journal} {\bibinfo  {journal}
  {Journal of multivariate analysis}\ }\textbf {\bibinfo {volume} {3}},\
  \bibinfo {pages} {337--394} (\bibinfo {year} {1973})}\BibitemShut {NoStop}%
\bibitem [{\citenamefont {Wilde}(2013)}]{wilde2013quantum}%
  \BibitemOpen
  \bibfield  {author} {\bibinfo {author} {\bibfnamefont {Mark~M}\ \bibnamefont
  {Wilde}},\ }\href@noop {} {\emph {\bibinfo {title} {Quantum information
  theory}}}\ (\bibinfo  {publisher} {Cambridge University Press},\ \bibinfo
  {year} {2013})\BibitemShut {NoStop}%
\bibitem [{\citenamefont {Avron}\ and\ \citenamefont
  {Elgart}(1999)}]{avron1999adiabatic}%
  \BibitemOpen
  \bibfield  {author} {\bibinfo {author} {\bibfnamefont {Joseph~E}\
  \bibnamefont {Avron}}\ and\ \bibinfo {author} {\bibfnamefont {Alexander}\
  \bibnamefont {Elgart}},\ }\bibfield  {title} {\enquote {\bibinfo {title}
  {Adiabatic theorem without a gap condition},}\ }\href@noop {} {\bibfield
  {journal} {\bibinfo  {journal} {Communications in mathematical physics}\
  }\textbf {\bibinfo {volume} {203}},\ \bibinfo {pages} {445--463} (\bibinfo
  {year} {1999})}\BibitemShut {NoStop}%
\bibitem [{\citenamefont {Teufel}(2001)}]{teufel2001note}%
  \BibitemOpen
  \bibfield  {author} {\bibinfo {author} {\bibfnamefont {Stefan}\ \bibnamefont
  {Teufel}},\ }\bibfield  {title} {\enquote {\bibinfo {title} {A note on the
  adiabatic theorem without gap condition},}\ }\href {\doibase
  10.1023/A:1014556511004} {\bibfield  {journal} {\bibinfo  {journal} {Letters
  in Mathematical Physics}\ }\textbf {\bibinfo {volume} {58}},\ \bibinfo
  {pages} {261--266} (\bibinfo {year} {2001})}\BibitemShut {NoStop}%
\bibitem [{\citenamefont {B\"urkle}\ and\ \citenamefont
  {Anglin}(2020)}]{burkle2020probabilistic}%
  \BibitemOpen
  \bibfield  {author} {\bibinfo {author} {\bibfnamefont {R.}~\bibnamefont
  {B\"urkle}}\ and\ \bibinfo {author} {\bibfnamefont {J.~R.}\ \bibnamefont
  {Anglin}},\ }\bibfield  {title} {\enquote {\bibinfo {title} {Probabilistic
  hysteresis in an isolated quantum system: The microscopic onset of
  irreversibility from a quantum perspective},}\ }\href {\doibase
  10.1103/PhysRevA.101.042110} {\bibfield  {journal} {\bibinfo  {journal}
  {Phys. Rev. A}\ }\textbf {\bibinfo {volume} {101}},\ \bibinfo {pages}
  {042110} (\bibinfo {year} {2020})}\BibitemShut {NoStop}%
\bibitem [{\citenamefont {Abou-Salem}\ and\ \citenamefont
  {Fr{\"o}hlich}(2005)}]{abou-salem2005adiabatic}%
  \BibitemOpen
  \bibfield  {author} {\bibinfo {author} {\bibfnamefont {Walid~K.}\
  \bibnamefont {Abou-Salem}}\ and\ \bibinfo {author} {\bibfnamefont {J{\"u}rg}\
  \bibnamefont {Fr{\"o}hlich}},\ }\bibfield  {title} {\enquote {\bibinfo
  {title} {Adiabatic theorems and reversible isothermal processes},}\ }\href
  {\doibase 10.1007/s11005-005-4838-1} {\bibfield  {journal} {\bibinfo
  {journal} {Letters in Mathematical Physics}\ }\textbf {\bibinfo {volume}
  {72}},\ \bibinfo {pages} {153--163} (\bibinfo {year} {2005})}\BibitemShut
  {NoStop}%
\bibitem [{\citenamefont {Abou-Salem}\ and\ \citenamefont
  {Fr{\"o}hlich}(2007)}]{abou-salem2007status}%
  \BibitemOpen
  \bibfield  {author} {\bibinfo {author} {\bibfnamefont {Walid~K.}\
  \bibnamefont {Abou-Salem}}\ and\ \bibinfo {author} {\bibfnamefont {J{\"u}rg}\
  \bibnamefont {Fr{\"o}hlich}},\ }\bibfield  {title} {\enquote {\bibinfo
  {title} {Status of the fundamental laws of thermodynamics},}\ }\href
  {\doibase 10.1007/s10955-006-9222-8} {\bibfield  {journal} {\bibinfo
  {journal} {Journal of Statistical Physics}\ }\textbf {\bibinfo {volume}
  {126}},\ \bibinfo {pages} {1045--1068} (\bibinfo {year} {2007})}\BibitemShut
  {NoStop}%
\bibitem [{\citenamefont {Jaksic}\ and\ \citenamefont
  {Pillet}(2014)}]{jaksic2014note}%
  \BibitemOpen
  \bibfield  {author} {\bibinfo {author} {\bibfnamefont {Vojkan}\ \bibnamefont
  {Jaksic}}\ and\ \bibinfo {author} {\bibfnamefont {Claude-Alain}\ \bibnamefont
  {Pillet}},\ }\bibfield  {title} {\enquote {\bibinfo {title} {A note on the
  landauer principle in quantum statistical mechanics},}\ }\href {\doibase
  10.1063/1.4884475} {\bibfield  {journal} {\bibinfo  {journal} {Journal of
  Mathematical Physics}\ }\textbf {\bibinfo {volume} {55}},\ \bibinfo {pages}
  {075210} (\bibinfo {year} {2014})}\BibitemShut {NoStop}%
\bibitem [{\citenamefont {Bachmann}\ \emph {et~al.}(2017)\citenamefont
  {Bachmann}, \citenamefont {De~Roeck},\ and\ \citenamefont
  {Fraas}}]{bachmann2016adiabatic}%
  \BibitemOpen
  \bibfield  {author} {\bibinfo {author} {\bibfnamefont {Sven}\ \bibnamefont
  {Bachmann}}, \bibinfo {author} {\bibfnamefont {Wojciech}\ \bibnamefont
  {De~Roeck}}, \ and\ \bibinfo {author} {\bibfnamefont {Martin}\ \bibnamefont
  {Fraas}},\ }\bibfield  {title} {\enquote {\bibinfo {title} {The adiabatic
  theorem for many-body quantum systems},}\ }\href@noop {} {\bibfield
  {journal} {\bibinfo  {journal} {Phys. Rev. Lett.}\ }\textbf {\bibinfo
  {volume} {119}},\ \bibinfo {pages} {060201} (\bibinfo {year}
  {2017})}\BibitemShut {NoStop}%
\bibitem [{\citenamefont {Venuti}\ \emph {et~al.}(2016)\citenamefont {Venuti},
  \citenamefont {Albash}, \citenamefont {Lidar},\ and\ \citenamefont
  {Zanardi}}]{venuti2016adiabaticity}%
  \BibitemOpen
  \bibfield  {author} {\bibinfo {author} {\bibfnamefont {Lorenzo~Campos}\
  \bibnamefont {Venuti}}, \bibinfo {author} {\bibfnamefont {Tameem}\
  \bibnamefont {Albash}}, \bibinfo {author} {\bibfnamefont {Daniel~A.}\
  \bibnamefont {Lidar}}, \ and\ \bibinfo {author} {\bibfnamefont {Paolo}\
  \bibnamefont {Zanardi}},\ }\bibfield  {title} {\enquote {\bibinfo {title}
  {Adiabaticity in open quantum systems},}\ }\href {\doibase
  10.1103/PhysRevA.93.032118} {\bibfield  {journal} {\bibinfo  {journal} {Phys.
  Rev. A}\ }\textbf {\bibinfo {volume} {93}},\ \bibinfo {pages} {032118}
  (\bibinfo {year} {2016})}\BibitemShut {NoStop}%
\bibitem [{\citenamefont {Benoist}\ \emph {et~al.}(2016)\citenamefont
  {Benoist}, \citenamefont {Fraas}, \citenamefont {Jaksic},\ and\ \citenamefont
  {Pillet}}]{benoist2016full}%
  \BibitemOpen
  \bibfield  {author} {\bibinfo {author} {\bibfnamefont {Tristan}\ \bibnamefont
  {Benoist}}, \bibinfo {author} {\bibfnamefont {Martin}\ \bibnamefont {Fraas}},
  \bibinfo {author} {\bibfnamefont {Vojkan}\ \bibnamefont {Jaksic}}, \ and\
  \bibinfo {author} {\bibfnamefont {Claude-Alain}\ \bibnamefont {Pillet}},\
  }\bibfield  {title} {\enquote {\bibinfo {title} {Full statistics of erasure
  processes: Isothermal adiabatic theory and a statistical landauer
  principle},}\ }\href {https://arxiv.org/abs/1602.00051} {\bibfield  {journal}
  {\bibinfo  {journal} {arXiv:1602.00051}\ } (\bibinfo {year}
  {2016})}\BibitemShut {NoStop}%
\bibitem [{\citenamefont {Teufel}(2019)}]{teufel2019non-equilibrium}%
  \BibitemOpen
  \bibfield  {author} {\bibinfo {author} {\bibfnamefont {Stefan}\ \bibnamefont
  {Teufel}},\ }\bibfield  {title} {\enquote {\bibinfo {title} {Non-equilibrium
  almost-stationary states and linear response for gapped quantum systems},}\
  }\href@noop {} {\bibfield  {journal} {\bibinfo  {journal} {Communications in
  Mathematical Physics}\ ,\ \bibinfo {pages} {1--33}} (\bibinfo {year}
  {2019})}\BibitemShut {NoStop}%
\end{thebibliography}%

\clearpage



\renewcommand{\theequation}{S\arabic{equation}}
\setcounter{equation}{0}

\begin{widetext}

\section{{\large Supplementary material} \label{sec: supplement}}

\subsection{Properties of $\mu_s$ and $\nu_s$}

Here we prove a Lemma about $\mu_s$ and $\nu_s$ required for the proof of the finite temperature adiabatic theorem. We introduce a shorthand notation
\be
\Delta_{mn}(s)\equiv E^m_s-E^n_s.
\ee
We assume that at a given $s$  the spectrum is ordered:
\be\label{ordering}
\Delta_{mn}(s)\geq 0 \qquad {\rm for} \qquad m>n.
\ee

Let us show that $\mu_s$ and $\nu_s$ defined respectively by eqs. \eqref{mu} and \eqref{nu} of the main text, satisfy the following\\

\smallskip

\noindent {\bf Lemma:}
\be\label{mu general}
\frac{1}{\mu( s)}=\max_{1\leq n<m \leq d}\left|\frac{\Delta_{mn}(0)}{\Delta_{mn}( s)}\right|
\ee
and
\be\label{nu general}
\nu( s)=\max_{1\leq n<m\leq d}\left|\frac{\partial_ s\Delta_{mn}( s)}{\Delta_{mn}( s)}\right|.
\ee

\noindent {\bf Proof:}~~ Consider an arbitrary set of real numbers $A_n,$ $n=1,2,\dots,d$ and introduce
\be\label{lemma1}
q=\max_{m,n}\left|\frac{A_{m}-A_{n}}{E^m_s-E^n_s}\right|.
\ee
We are going to prove that in fact
\be\label{lemma equality}
q=\max_{k}\left|\frac{A_{k+1}-A_{k}}{E^{k+1}_s-E^k_s}\right|.
\ee
This equality entails eq. \eqref{mu general} for $A_n=E_0^n$  and  eq. \eqref{nu general} for $A_n=\partial_s E^n_s$.

To prove eq. \eqref{lemma equality}, we start from an obvious observation that
\be
q\geq\max_{k}\left|\frac{A_{k+1}-A_{k}}{E^{k+1}_s-E^k_s}\right|.
\ee
Let us show that, in fact, the strict inequality is impossible. To this end we assume the opposite, i.e. that
\be\label{opposite}
\left|\frac{A_{k+1}-A_{k}}{E^{k+1}_s-E^k_s}\right|< q\qquad \forall k.
\ee
Then for any $m>n$ we obtain
\be\label{lemma2}
|A_{m}-A_{n}|=\left|\sum_{k=n}^{m-1}(A_{k+1}-A_{k})\right|\leq\sum_{k=n}^{m-1}|A_{k+1}-A_{k}|<q\sum_{k=n}^{m-1}(E^{k+1}_s-E^k_s)=q(E^m_s-E^n_s),
\ee
where the ordering of energies, \eqref{ordering}, is used to get rid of the modulus. Eq. \eqref{lemma2} is  inconsistent with  eq. (\ref{opposite}). Thus the equality \eqref{lemma equality} is true, q.e.d.

\subsection{Proof of the finite temperature adiabatic theorem}


Here we prove the bound \eqref{main result} from the main text.

We introduce
\be
\sigma_t=U_{\omega t}^\dag\rho_{t}U_{\omega t},
\ee
which evolves according to
\be
\partial_t \sigma_t=-i[\widetilde{H}_{\omega t}+\omega V_{\omega t},\sigma_{t}],\qquad \sigma_{0}=e^{-\beta  H_{0}}/Z_0.
\ee
We denote by dot the derivative of a function with respect to its argument $s=\omega t$. For example,  $\dot{U}_{ s}=\partial_ s U_{ s}$, $\dot{U}_{\omega t}=\partial_ s U_{s}|_{s=\omega t}$  but $\partial_t U_{\omega t}=\omega\dot{U}_{\omega t}$.

We first estimate the quantity
\be
D \equiv 1-\tr(\sqrt{\theta_t^\beta}\sqrt{\rho_{t}})=1-\tr(\sqrt{\rho_{0}}\sqrt{\sigma_{t}}).
\ee

Note that
\be
\partial_t \sqrt{\sigma_t}=-i[\widetilde{H}_{\omega t}+\omega V_{\omega t},\sqrt{\sigma_t}].
\ee
Therefore
\begin{align}
\nonumber
\partial_t D &  = -\tr(\sqrt{\rho_{0}}\, \partial_t \sqrt{\sigma_{t}})=i \, \frac{1}{\sqrt{Z_0}}\tr(e^{-\beta  H_0/2}\, [\widetilde{H}_{\omega t}+\omega V_{\omega t},\sqrt{\sigma_t}])=
i\omega \, \frac{1}{\sqrt{Z_0}}\tr([e^{-\beta  H_0/2}, V_{\omega t}] \sqrt{\sigma_t})\\[8 pt]
\nonumber
 & =  \frac{i\omega}{\sqrt{Z_0}}\,\sum_{n,\, k} (e^{-\frac{\beta}{2}E^n_0}-e^{-\frac{\beta}{2}E^k_0}) \, \langle n| V_{\omega t} |k \rangle \,  \langle k | \sqrt{\sigma_t} |n\rangle
 \\[8 pt]
 & =
 \frac{i\beta \omega}{2\sqrt{Z_0}}  \,\sum_{n,\, k} f_{nk}(\omega t) \Delta_{nk}(\omega t) \, \langle n| V_{\omega t} |k \rangle \,  \langle k | \sqrt{\sigma_t} |n\rangle,
 \label{dD}
\end{align}
where $|n \rangle=|\Phi_{0}^n \rangle$, $|k \rangle=|\Phi_{0}^k \rangle$ are eigenstates of $H_0$ and, consequently, of $\widetilde{H}_{\omega t}$  for arbitrary $t$, $E^n_0, E^k_0$ are  eigenenergies of $H_{0}$, and
\be\label{fnk Delta nk definitions}
f_{nk}(\omega t)\equiv\frac{e^{-\frac{\beta}{2}E^n_0}-e^{-\frac{\beta}{2}E^k_0}}{\beta\Delta_{nk}(\omega t)/2}~~{\rm for}~~n\neq k,~~~f_{nn}=0.
\ee
Note that we will occasionally drop an argument of the function  $f_{nk}$ when this does not lead to ambiguities.


We notice that
\be\label{first route}
(E^n_{\omega t}-E^k_{\omega t}) \,  \langle k | \sqrt{\sigma_t} |n\rangle=-\langle k | [\widetilde{H}_{\omega t},\sqrt{\sigma_t}] |n\rangle = -i \langle k | \partial_t\sqrt{\sigma_t} |n\rangle+
\omega \langle k | [V_{\omega t},\sqrt{\sigma_t}] |n\rangle.
\ee
Substituting this expression to eq. \eqref{dD} and integrating it over time one obtains
\be\label{D}
D=\frac{\beta\omega}{2\sqrt{Z_0}} \sum_{n,\, k} \int_{0}^{t}f_{nk}(\omega t')\langle n| V_{\omega t'} |k \rangle \left(\langle k|\partial_{t'}\sqrt{\sigma_{t'}} |n\rangle +i\omega\langle k | [V_{\omega t'},\sqrt{\sigma_{t'}}]|n\rangle  \right)dt'.
\ee
Integrating (\ref{D}) by parts one gets
\begin{eqnarray}
\nonumber
D&=&\frac{\beta\omega}{2\sqrt{Z_0}}  \sum_{n,\, k}\Bigg(f_{nk}(\omega t)\langle n| V_{\omega t} |k \rangle \langle k|\sqrt{\sigma_{t}} |n\rangle-\omega\int_{0}^{t}f_{nk}(\omega t')\langle n| \dot{V}_{\omega t'} |k \rangle \langle k|\sqrt{\sigma_{t'}} |n\rangle dt'
\\
\nonumber
&+&\omega\int_{0}^{t}f_{nk}(\omega t')\langle n| V_{\omega t'} |k \rangle \frac{\dot{\Delta}_{nk}(\omega t')}{\Delta_{nk}(\omega t')}\langle k|\sqrt{\sigma_{t'}} |n\rangle dt'+i\omega\int_{0}^{t}f_{nk}(\omega t')\langle n| V_{\omega t'} |k \rangle \langle k | [V_{\omega t'},\sqrt{\sigma_{t'}}]|n\rangle dt'\Bigg)
\\
&=&\frac{\beta\omega}{2\sqrt{Z_0}} (K_1+K_2+K_3+K_4),
\end{eqnarray}
where we us $\dot{f}_{nk}(\omega t)=-f_{nk}(\omega t)\dot{\Delta}_{nk}(\omega t)/\Delta_{nk}(\omega t)$ and
\begin{eqnarray}
 K_1&=&\sum_{n,\, k}f_{nk}(\omega t)\langle n| V_{\omega t} |k \rangle \langle k|\sqrt{\sigma_{t}} |n\rangle \label{K1}
 \\
  K_2&=&-\omega\sum_{n,\, k}\int_{0}^{t}f_{nk}(\omega t')\langle n| \dot{V}_{\omega t'} |k \rangle \langle k|\sqrt{\sigma_{t'}} |n\rangle dt'
  \\
 K_3&=&\omega\sum_{n,\, k}\int_{0}^{t}f_{nk}(\omega t')\langle n| V_{\omega t'} |k \rangle \frac{\dot{\Delta}_{nk}(\omega t')}{\Delta_{nk}(\omega t')}\langle k|\sqrt{\sigma_{t'}} |n\rangle dt'
 \\
K_4&=&i\omega\sum_{n,\, k}\int_{0}^{t}f_{nk}(\omega t')\langle n| V_{\omega t'} |k \rangle \langle k | [V_{\omega t'},\sqrt{\sigma_{t'}}]|n\rangle dt'
\end{eqnarray}

Obviously,
\be\label{D<K+K+K+K}
|D|\leq \frac{\beta\omega}{2\sqrt{Z_0}} (|K_1|+|K_2|+|K_3|+|K_4|).
\ee


Let us estimate $|K_1|$:
\begin{align}\label{K11}
|K_1|\leq
\left( \sum_{n,\, k} f_{nk}^2 \, \langle n| V_{\omega t} |k \rangle\,\langle k| V_{\omega t} |n \rangle \right)^{1/2}
\left( \sum_{n,\, k} \langle k | \sqrt{\sigma_t} |n\rangle \langle n | \sqrt{\sigma_t} |k\rangle \right)^{1/2}.
\end{align}
The term in the second bracket reads
\be
\sum_{n,\, k} \langle k | \sqrt{\sigma_t} |n\rangle \langle n | \sqrt{\sigma_t} |k\rangle=\tr \sigma_t=\tr \rho_0=1.
\ee
To estimate the first term we need to estimate $f_{nk}^2$:
\be\label{fnk estimate}
f_{nk}^2(\omega t)=\left(\frac{e^{-\frac{\beta}{2}E^n_0}-e^{-\frac{\beta}{2}E^k_0}}{\beta(E^n_0-E^k_0)/2}\frac{\Delta_{nk}(0)}{\Delta_{nk}(\omega t')}\right)^2\leq\left(\frac{e^{-\frac{\beta}{2}E^n_0}-e^{-\frac{\beta}{2}E^k_0}}{\beta(E^n_0-E^k_0)/2}\right)^2\frac{1}{\mu^2(\omega t)},
\ee
where eq. \eqref{mu general} is used to establish the inequality.
Further, by the Lagrange's Mean Value Theorem there exists $a\in(0,1)$ such that
\be\label{fnk estimate 2}
\left(\frac{e^{-\frac{\beta}{2}E^n_0}-e^{-\frac{\beta}{2}E^k_0}}{\beta(E^n_0-E^k_0)/2}\right)^2=e^{-\beta(a E^n_0+(1-a)E^k_0)}\leq
e^{-\beta\min \{E^n_0,E^k_0\}}\leq e^{-\beta E^n_0}+e^{-\beta E^k_0},\quad n\neq k.
\ee
Combining inequalities \eqref{fnk estimate} and \eqref{fnk estimate 2} and extending them to the trivial case $n=k$ (where $f_{nn}=0$ by definition  \eqref{fnk Delta nk definitions}) we get
\be
f_{nk}^2(\omega t)\leq \frac{1}{\mu^2(\omega t)}\,\left(e^{-\beta E^n_0}+e^{-\beta E^k_0}\right).
\ee
We use this bound to proceed further:
\begin{eqnarray}
\label{estimate 1}
\nonumber
&&\sum_{n,\, k} f_{nk}^2 \, \langle n| V_{\omega t} |k \rangle\,\langle k| V_{\omega t} |n \rangle
\\
\nonumber
&\leq&\frac{1}{\mu^2(\omega t)}\sum_{n,\, k}\left(e^{-\beta E^n_0} \langle n| V_{\omega t} |k \rangle\,\langle k| V_{\omega t} |n \rangle+e^{-\beta E^k_0} \langle n| V_{\omega t} |k \rangle\,\langle k| V_{\omega t} |n \rangle\right)
\\
\nonumber
&=& \frac{1}{\mu^2(\omega t)}\sum_{n,\, k}\Big(\langle n| V_{\omega t} |k \rangle\,\langle k|   V_{\omega t} \, e^{-\beta H_0}|n \rangle+\langle n| V_{\omega t} |k \rangle\,\langle k| e^{-\beta H_0}\, V_{\omega t} |n \rangle\Big)
\\
&=&2\frac{1}{\mu^2(\omega t)}\tr (V_{\omega t}^2 e^{-\beta H_0})
\end{eqnarray}
Next use the  inequality
\be
|\tr AB|\leqslant\|A\|\tr B,
\ee
valid for any  $B>0$ and diagonalisable $A$ (be reminded that $\|...\|$ stands for the operator norm), to obtain
\be
\tr V_{\omega t}^2 e^{-\beta H_0}\leq \|V_{\omega t}^2\|\,\tr \, e^{-\beta H_0}.
\ee
Finally
\be\label{K1}
|K_1| \leq  \sqrt{2Z_0}\frac{1}{\mu_{\omega t}}\|V_{\omega t}\|.
\ee

$K_2$ can be bounded in an analogous way:
\be
\left|\sum_{n,\, k}f_{nk}(\omega t')\langle n| \dot{V}_{\omega t'} |k \rangle \langle k|\sqrt{\sigma_{t'}} |n\rangle\right|\leq\sqrt{2Z_0}\frac{1}{\mu_{\omega t'}}\|\dot{V}_{\omega t'}\|
\ee
and
\be\label{K2}
|K_2|\leq\omega\sqrt{2Z_0}\int^{t}_0\frac{1}{\mu_{\omega t'}}\|\dot{V}_{\omega t'}\|dt'=\sqrt{2Z_0}\int^{\omega t}_0\frac{1}{\mu_{s'}}\|\dot{V}_{ s'}\|d s'.
\ee
Let us estimate $|K_3|$:
\begin{align}
|K_3|\leq
\int^t_0\left( \sum_{n,\, k} f_{nk}^2 \left(\frac{\dot{\Delta}_{nk}(\omega t')}{\Delta_{nk}(\omega t')}\right)^2 \langle n| V_{\omega t'} |k \rangle\,\langle k| V_{\omega t} |n \rangle \right)^{1/2}
\left( \sum_{n,\, k} \langle k | \sqrt{\sigma_{t'}} |n\rangle \langle n | \sqrt{\sigma_{t'}} |k\rangle \right)^{1/2}dt',
\end{align}
\be
f_{nk}^2(\omega t')\left(\frac{\dot{\Delta}_{nk}(\omega t')}{\Delta_{nk}(\omega t')}\right)^2=\left(\frac{e^{-\frac{\beta}{2}E^n_0}-e^{-\frac{\beta}{2}E^k_0}}{\beta(E^n_0-E^k_0)/2}\frac{\Delta_{nk}(0)\dot{\Delta}_{nk}(\omega t')}{\Delta^2_{nk}(\omega t')}\right)^2\leq\frac{\nu^2(\omega t')}{\mu^2(\omega t')}(e^{-\beta E^n_0}+e^{-\beta E^k_0}),
\ee
where eqs. \eqref{mu general}, \eqref{nu general} are used to establish the inequality. Thus we obtain
\be\label{K3}
|K_3|\leq\sqrt{2Z_0}\int^{\omega t}_0\frac{\nu_{s'}}{\mu_{s'}}\|V_{ s'}\|d s'
\ee
Finally, let us estimate  $K_4$:
\begin{align}
|K_4|\leq \omega\int_{0}^{t}dt'
\left( \sum_{n,\, k} f_{nk}^2 \, \langle n| V_{\omega t'} |k \rangle\,\langle k| V_{\omega t'} |n \rangle \right)^{1/2}
\left( \sum_{n,\, k} \langle k | [V_{\omega t'},\sqrt{\sigma_{t'}}] |n\rangle \langle n | [\sqrt{\sigma_{t'}},V_{\omega t'}] |k\rangle \right)^{1/2}.
\end{align}
The term in the first bracket has been already bounded, see eq. \eqref{estimate 1}.
The term in the second bracket reads
\be
\tr [V_{\omega t'},\sqrt{\sigma_{t'}}]\,[\sqrt{\sigma_{t'}},V_{\omega t'}]=2 \, \tr V_{\omega t'}^2 \sigma_{t'} -2 \, \tr \left( \sigma_{t'}^{1/4} V_{\omega t'}  \sigma_{t'}^{1/4}\right)^2\leq 2 \|V_{\omega t'}^2\| \, \tr \sigma_t=2 \|V_{\omega t'}\|^2,
\ee
and we get
\be\label{K44}
|K_4| \leq  2\sqrt{Z_0} \,\omega\int_{0}^{t}\frac{1}{\mu_{\omega t'}}\|V_{\omega t'}\|^2dt'=2\sqrt{Z_0}\,\int_{0}^{\omega t}\frac{1}{\mu_{s'}}\|V_{ s'}\|^2d s'.
\ee

Finally we collect all pieces \eqref{K1}--\eqref{K44} together and bound $D$ according to eq. \eqref{D<K+K+K+K}:
\begin{eqnarray}\label{D fin}
D&\leq& \frac{\omega\beta}{\sqrt{2}} \,\left(\frac{1}{\mu_{\omega t}}\|V_{\omega t}\|+\int_{0}^{\omega t}\frac{1}{\mu(t')}\|\partial_{t'}V_{t'}\|dt'+\int_{0}^{\omega t}\frac{\nu(t')}{\mu(t')}\|V_{t'}\|dt'+\sqrt2 \, \int_{0}^{\omega t}\frac{1}{\mu(t')}\|V_{t'}\|^2dt'\right).
\end{eqnarray}

The last thing we need to do is to connect $D$ with the trace distance $ D_{\rm tr}\left(\rho_t,\theta_t^\beta\right)$. This can be done thanks to the inequality proven in \cite{holevo1972quasiequivalence} which reads

\be\label{Holevo inequality}
 D_{\rm tr}\left(\rho_1,\rho_2\right)\leq \sqrt{1-\left(\tr \sqrt\rho_1\sqrt\rho_2\right)^2}\leq \sqrt{2 D}.
\ee
Eq. \eqref{main result} of the main text  follows from eqs. \eqref{D fin} and \eqref{Holevo inequality}, q.e.d.

As was mentioned in the main text, the presence of operator norms in the final result makes the bound inapplicable in the cases where $V$ or $\dot{V}$ are unbounded operators. In fact, the operator norms are superficial for estimating $K_1$, $K_2$ and $K_3$ above and can be substituted by thermal averages with respect to the initial Gibbs state. For $K_1$ this can be seen from the eq. \eqref{estimate 1}, and analogously for $K_2$ and $K_3$. However, we were not able to avoid the operator norm when estimating~$K_4$.

\subsection{Spin moved around a wire: pure state adiabaticity}

Here we derive a condition for pure state adiabaticity in the electron-spin system (see \eqref{H wire},\eqref{H Se} of the main text) under the assumption of periodic boundary conditions for electron wave functions in the $z$-direction. To be specific, we choose
\be
S=\frac12.
\ee

Using eq. \eqref{current} of the main text, we rewrite the Hamiltonian as
\be\label{Hamiltonian wire suppl}
H_\alpha=H_e+ \gamma \, e^{-i\alpha S_z} S_y  \, e^{i\alpha S_z},
\ee
where
\be\label{gamma}
\gamma = -\frac{\mu_{\rm magn}}{2\pi r} \frac{e\rho A }{m_e N} \,P_e.
\ee
The total momentum of electrons, $P_e$, commutes with $H_\alpha$, therefore we treat it as a $c$-number.
We initialize the system in an eigenstate  of $H_\alpha$,
\be
\Psi_0=|{\rm electrons} \rangle \otimes  \psi_0,
\ee
where $|{\rm electrons} \rangle$ is an eigenstate of $H_e$ and $P_e$, while $\psi_0$ is an eigenstate of $S_y $,
\be
S_y  \psi_0 =\frac12 \psi_0.
\ee

The time-dependent many-body wave function $\Psi_t$ satisfies the Schr\"odinger equation
\be
i\partial_t \Psi_t=H_{\omega t} \Psi_t.
\ee
It is easy to see that $\Psi_t$  factors as follows:
\be\label{Psi_t factorized}
\Psi_t =\left( e^{-i H_e t} |{\rm electrons} \rangle \right)\otimes \psi_t,
\ee
where $\psi_t$ satisfies the Schr\"odinger equation with the effective spin Hamiltonian $H^{Se}_{\omega t} $,
\be\label{effective SE suppl}
i\partial_t \psi_t = H^{Se}_{\omega t} \psi_t,\qquad H^{Se}_{\omega t}=\gamma \, e^{-i\omega t S_z} S_y  \, e^{i\omega t S_z}.
\ee

The figure of merit of the pure state adiabaticity is the adiabatic fidelity between the dynamical wave function $\Psi_t$ and the instantaneous eigenfunction $\Phi_\alpha$ of the Hamiltonian \eqref{Hamiltonian wire suppl}:
\be
{\cal F}_t\equiv |\langle \Phi_{\alpha=\omega t}| \Psi_t\rangle|^2.
\ee
From \eqref{Psi_t factorized} one immediately obtains that ${\cal F}_t$ is given by
\be
{\cal F}_t= |\langle \varphi_{\alpha=\omega t}|\psi_t\rangle|^2,
\ee
where $\varphi_\alpha$ is the eigenstate of $H^{Se}_\alpha$ satisfying $\varphi_0=\psi_0$.

The dynamics of $\psi_t$ can be easily inferred from eq. \eqref{effective SE suppl} by transformation to the rotating frame. As a result one obtains
\begin{equation}\label{fidelity}
    1 -  {\cal F}_t = \left. \frac{\omega^2}{\omega^2 + \gamma^2} \, \sin^2\left( \frac\alpha2 \sqrt{1 + \gamma^2 / \omega^2} \right) \right|_{\alpha=\omega t}.
\end{equation}

We say that the adiabaticity is maintained up to some target $\alpha$ with the accuracy $\varepsilon$ if for $t\leq\alpha/\omega$
\be
1 -  {\cal F}_t\leq \varepsilon.
\ee
Let us assume that the target $\alpha$ is greater than $\pi$. Then the sine squared in eq. \eqref{fidelity} will become equal to unity somewhere on the way to the target $\alpha$. Taking this into account, we conclude from eq. \eqref{fidelity} that the maximal driving rate $\omega_{\varepsilon}$ that allows to maintain adiabaticity with the accuracy $\varepsilon$ is given by
\be\label{adiabatic driving}
\omega_{\varepsilon} = \gamma\,\sqrt{\frac{\varepsilon}{1-\varepsilon}}.
\ee
Since $P_e\sim p_F \sqrt{N}$ in the majority of states in the Gibbs ensemble, it follows from eqs. \eqref{gamma} and \eqref{adiabatic driving} that for these states
\be\label{sqrt scaling}
\omega_\varepsilon \sim 1/\sqrt N
\ee
in the thermodynamic limit $N\rightarrow \infty$, $\rho=\const$.

It should be stressed that the explicit solution of the dynamical problem presented here works only in the case when the total momentum of electrons in the wire is conserved, i.e. for periodic boundary conditions imposed on electron wave functions.  If this is not the case, e.g. for a long piece of wire with open ends, the dynamics of electrons and the spin are coupled. However, we see no reasons to expect that different boundary conditions can make the scaling of the driving rate with the system size more favorable for the pure state adiabaticity.

%
%
%

\end{widetext}

\end{document}